\documentclass[12pt]{article}
\usepackage{todonotes}
\newcommand\comments[1]{}%
\newcommand\newversion[1]{#1}%
\newcommand\oldversion[1]{}%
\newcommand\todof[1]{}%
\newcommand{\todonotesexplained}{}
\usepackage{imakeidx}
\makeindex[name=notations,title=Index of notations,columns=1,options=-s indexstyle.ist]
\usepackage{ulem}
\usepackage{float}
\usepackage{placeins}
\usepackage{color}
\usepackage{xcolor}
\usepackage{enumerate}

\usepackage[latin1]{inputenc}
\usepackage{amsthm,amsmath}
\usepackage{amsfonts}
\usepackage{booktabs}
\usepackage{tikz}
\usepackage{mathrsfs}                  
\usepackage{dsfont}
\usepackage{natbib} 

\usepackage{framed}
\usepackage[colorlinks,citecolor=gray,urlcolor=gray]{hyperref}
\usepackage{graphics}
\usepackage{times}
\newcommand\total{\mathrm{t}}
\newcommand\random{\eta}

\newcommand\mon{m}
\newcommand\indiv{k}
\newcommand\Mon{M}

\newcommand\paramy{\mathbf{y}}
\newcommand\paramw{\mathbf{w}}
\newcommand\paramx{\mathbf{x}}
\newcommand\paramz{\mathbf{z}}
\DeclareMathOperator*{\argmin}{argmin}

\newcommand{\urf}{\mathrm{R}}
\newcommand{\ur}{\mathrm{r}}
\newcommand\rate{\ur}
\newcommand\rotationbias{b}
\newcommand{\direct}{\mathrm{direct}}
\newcommand{\ak}{\mathrm{AK}}

\newcommand{\mis}{\mathrm{mis}}
\newcommand\rcind{\mathrm{r.c.}}
\newcommand\rc{\text{r.c.}}
\newcommand\paramzc[1][]{\mathbf{z}^{\rc #1}}
\newcommand\sample{S}
\newcommand{\misi}{g}

\newcommand{\transp}[1]{\left.#1\right.^{\!\mathrm{T}}}
\newcommand{\partransp}[1]{\left(#1\right)^{\!\mathrm{T}}}
\newcommand{\numofhinrg}{n}
\newcommand{\numofhinpop}{H}
\newcommand{\rotationgroup}{\mathrm{Clu}}
\newcommand{\household}{h}
\newcommand{\status}{e}
\newcommand{\Xmatrix}{X}

\begin{document}

\begin{center}
{\fontsize{16pt}{20pt}\selectfont {\textbf {An Evaluation of Design-based Properties of Different Composite Estimators}}}\\[.5cm]

{\fontsize{12pt}{14pt}{\textbf {Daniel Bonn\'ery$^{\mathrm{1}}$,  Yang Cheng$^{\mathrm{2}}$, Partha Lahiri$^{\mathrm{3}}$}}}\\[.5cm]

$^{\mathrm{1}}$ JPSM and US Census Bureau Research Associate, $^{\mathrm{2}}$ US Census Bureau,
$^{\mathrm{3}}$ JPSM, University of Maryland

$^{\mathrm{1,3}}$JPSM, 1218 LeFrak Hall,  7251 Preinkert Dr., College Park, MD 20742, USA,
$^{\mathrm{2}}$4600 Silver Hill Rd, Washington, MD 20233, United States\\[.5cm]

\end{center}

\begin{framed}Disclaimer:
Any views expressed are those of the authors and not necessarily those of the U.S. Census Bureau.\end{framed}
\todonotesexplained

\section*{Abstract}

For the last several decades, the US Census Bureau has been using the AK composite estimation method to produce statistics on employment from the Current Population Survey (CPS) data. The CPS uses a rotating design and AK estimators are linear combinations of monthly survey weighted averages (called month-in-sample estimates) in each rotation groups. Denoting by $X$ the vector of month-in-sample estimates and by $\Sigma$ its design based variance, the coefficients of the linear combination were optimized by the Census Bureau after substituting $\Sigma$ by an estimate and under unrealistic stationarity assumptions. To show the limits of this approach, we compared the AK estimator with different competitors using three different synthetic populations that mimics the Current Population Survey (CPS) data and a simplified sample design that mimics the CPS design. In our simulation setup, empirically best estimators have larger mean square error than simple averages. In the real data analysis, the AK estimates are constantly below the survey-weighted estimates, indicating potential bias. Any attempt to improve on the estimated optimal estimator in either class would require a thorough investigation of the highly non-trivial problem of estimation of $\Sigma$ for a complex setting like the CPS (we did not entertain this problem in this paper).  A different approach is to use a variant of the regression composite estimator used by Statistics Canada. The regression composite estimator does not require estimation of $\Sigma$ and is less sensitive to the rotation group bias in our simulations. Our study demonstrates  that there is a great potential for improving the estimation of levels and month to month changes in the unemployment rates by using the regression composite estimator.

\subsection*{Keywords}
Calibration; estimated controls; longitudinal survey; labor force statistics.

\section{Introduction}

In repeated surveys, including rotating panel surveys, statistical data integration plays an important role in producing efficient estimators by extracting relevant information over time. 
Different ways to tackle changing samples have been proposed; see \cite{jones1980best}, \cite{yansaneh1998optimal}, \cite{bell2001comparison}, \cite{singh2001regression}, \cite{fuller2001regression} and others.  Such composite estimators typically improve on the standard direct survey-weighted  estimators in terms of mean squared error (MSE) and are commonly used by different government agencies for producing official labor force statistics. For example, to produce national employment and unemployment levels and rates, the U.S. Census Bureau uses the  AK composite estimation technique developed using the ideas given in \cite{gurney1965multivariate}.

Motivated by a Statistics Canada application, \cite{singh1995composite} introduced an ingenious idea for generating a composite estimator that can be computed using Statistics Canada's existing software for computing generalized regression estimates.  The key idea in \cite{singh1995composite} is to create a proxy (auxiliary) variable that uses information at the individual level as well as estimates at the population level from both previous and current periods. Using this proxy variable,  \cite{singh1995composite} obtained a composite estimator, referred to as Modified Regression 1 estimator (MR1) in the literature. However, \cite{Singh1997} noted that MR1 does not perform well in estimating changes in labor force statistics, which motivated them to propose a different composite estimator, called MR2, using a new proxy variable.  \cite{singh2001regression} generalized the idea of MR1 and MR2 estimators by suggesting a general set of proxy variables.

\cite{fuller2001regression} noted that the regression composite estimator proposed by  \cite{Singh1997} is subject to an undesirable drift problem, i.e., it may produce estimates that drift away from the real value suggested by the underlying model as time progresses and proposed an alternative regression composite method to rectify the drift problem.  Their method differs from the method of \cite{singh2001regression} in two directions.  First, the idea of rectifying the drift problem by a weighted combination of the two proxy variables used for MR1 and MR2 is new.
 Secondly, their final regression composite estimator involves estimation of the weight assigned to MR1 or MR2 control variable in the weighted combination --- this idea was not discussed in \cite{singh2001regression}.  In short, the Fuller-Rao regression composite estimator with estimated weight cannot be viewed as a special case of \cite{singh2001regression} and vice versa.

\cite{gambino2001regression} conducted an empirical study to evaluate the Fuller-Rao regression composite estimator, offered missing value treatment and listed several advantages (e.g, weighting procedure, consistency, efficiency gain, etc.) of the Fuller-Rao regression composite estimator over the AK estimator. Statistics Canada now uses the Fuller-Rao method for their official labor force statistics production. \cite{salonen2007regression} conducted an empirical study to compare the currently used Finnish labor force estimator with the Fuller-Rao's regression composite and other estimators.
  \cite{bell2001comparison}
 applied the generalized regression technique to improve on the Best Linear Unbiased Estimator (BLUE) based on a fixed window of time points and compared his estimator with the AK composite estimator
 of \cite{gurney1965multivariate} and the modified regression estimator of Singh et al. (1997), using data from the Australian Labour Force Survey. \cite{beaumont2005refinement} proposed a regression composite estimator with missing covariates defined using variables of interest from the previous month.
 
The main goal of this paper is to compare the design-based properties of the AK estimator with different rival estimators using the  CPS data.  To this end, we first expand the list of potential estimators by considering two new classes of composite estimators.  The first class includes the AK estimator as a member.  The second class generalizes the class of estimators considered earlier by \cite{yansaneh1998optimal} to incorporate multiple categories of employment status (e.g.,  employed, unemployed, and not in the labor force).  We obtain the best linear unbiased estimator (BLUE) for each class of estimators.  We call them the best AK estimator and multivariate BLUE, respectively.  As special cases of the multivariate BLUE, one can generate the univariate BLUE and the best AK estimators.  If the covariance matrix between two vectors of observations corresponding to any two different variables is a null matrix, then multivariate BLUE is identical to the univariate BLUE when the design matrix is the same for the variables. However, in general they are not identical when we do not have a block-diagonal covariance structure as is the case in our problem. 

     The optimal estimator for a given class of estimators, derived under given model and optimality condition cannot be used as it involves unknown model parameters (e.g., variances and covariances).  The AK estimator used by the Census Bureau is obtained from the optimal estimator when variances and covariances are substituted by  estimators justified under a rather strong stationary assumptions. We devise an evaluation study in order to assess the exact design-based properties of different composite estimators using the CPS data and CPS sample design. We demonstrate that the optimal estimator for a given model with estimated variances and covariances can perform poorly even when the modeling assumptions are valid. We included the multivariate BLUE with estimated variances and covariances for completeness of this research. While the multivariate BLUE performs the best under the model that generates it, as expected, it performed the worst (worst than the univariate BLUE with estimated variances and covariances) once we substitute estimated variances and covariances in the multivariate BLUE formula. Overall, we found that the Fuller-Rao estimator performed the best among all composite estimators considered in our study.              
 
In Section 2, we discuss the population and sample design. 
In Section 3, we review different classes of estimators and optimal estimator within each class.
In Section 4, we describe our evaluation study to assess the design-based properties of different estimators. 
In section 5, we report the CPS data  analysis. 
Some discussions and future research topics are given in Section 6.  We defer the proofs of relevant results and description of CPS design to the Appendix.  To facilitate reading of the paper, in the appendix we list all the notations used in the paper.
\section{Notations}
\subsection{Population}
Our theoretical framework uses three indices to identify three dimensions: $m$ for month, $k$ for individual and $e$ for an employment status category.  In this paper, we will consider three categories of employment status: employed, unemployed and not in the labor force.  The theory and methods developed in this paper, however, extend to more than 3 categories of employment status in a straightforward way.
Consider a sequence of finite populations of individuals $\left(U_\mon\right)_{\mon\in \left\{1\ldots M\right\}}$, where  $U_\mon$\index[notations]{Um@$U_\mon$ : population at month $\mon$}\index[notations]{U@$U=\bigcup_\mon^\Mon U_\mon$ : union over time of all the monthly populations }
refers to the finite population for month $\mon$.\index[notations]{m@$\mon$ : month index} Let $N$ denote the cardinality of $U=\bigcup_{m=1}^M U_m$.
Let $\paramy_{\mon,\indiv,\status}=1$ if the $\indiv$th individual belongs to $U_\mon$ and has $\status$th employment status and $\paramy_{\mon,\indiv,\status}=0$ otherwise, $\mon\in\{1,\cdots,\Mon\},\;\indiv\in\{1,\cdots,N\}, \; \status\in\{1,2,3\}.$  Because of our three dimensional data structure, we find it convenient to introduce arrays in developing our methodology and theory. 
\index[notations]{k@$\indiv$ : individual index, $\indiv=1,\ldots,N$}
Let  
$\paramy=[\paramy_{\mon,\indiv,\status}]_{\mon\in\{1,\ldots,\Mon\},\indiv\in \{1,\ldots,N\},e\in \{1,2,3\}}$ denote a three dimensional  $(\Mon,N,3)$-sized array.
\index[notations]{e@$\status$ : employment status index, 1: employed, 2: unemployed, 3: not in the labor force}\index[notations]{ym@$\paramx,\ \paramy,\ \paramz$ : 3-dimensional arrays of variables (auxiliary, study and endogenous, respectively) indexed by month, individual, and variable}
 We also define $\paramx$ 
 as a 3-dimensional array of auxiliary variables indexed by month, individual, and auxiliary variable, and an array $\paramz$, indexed the same way,
 that contains endogenous variables in the sense that $\paramz$ is a function of $\paramx$ and $\paramy$. 
 Any element of an array with ($\mon,\indiv$)-index satisfying $k\notin U_\mon$ is equal to 0 by convention.

\subsection{Notational conventions on arrays}
Given subsets $A$, $B$, $C$ of $\{1,\ldots,M\}$, $\{1,\ldots,N\}$, $\{1,2,3\}$, respectively (including the full set), we use the following notation  for sub-arrays: $\paramy_{A,B,C}=[\paramy_{a,b,c}]_{a\in A,b\in B, c\in C}$, and may replace $A$, $B$, or $C$ by ``.'' when $A=\{1,\ldots,M\}$, $B=\{1,\ldots,N\}$ or 
$C=\{1,2,3\}$, respectively: for example, $\paramy=\paramy_{.,.,.}$.
Let $\total_\paramy =\left[\sum_{\indiv\in U}\paramy_{\mon,\indiv,\status}\right]_{\mon\in\{1,\ldots,\Mon\},\status\in\{1,2,3\}}$ be the two dimensional $(\Mon,3)$-sized array  of population totals  indexed by month $\mon$ and employment status $\status$.\index[notations]{t@$\total_\paramy$ :  $(\Mon, 3)$-sized matrix indexed by month and employment status, $(\total_\paramy)_{(\mon,\status)}$ is the population count of individuals with status $\status$ in month $\mon$} 
We now show we can form a vector or matrix from an array. For a $p$-dimensional $(a_1,\ldots,a_p)$-sized array $A$, define $\vec{A}$ as the vector $\left(\vec{A}_1,\ldots,\vec{A}_{\prod_{l=1}^pa_l}\right)$, 
where $\forall (i_1,\ldots,i_p)\in \prod_{l=1}^p\{1,\ldots,a_l\}$, $\vec{A}_{1+\sum_{l=1}^p \left[\prod_{l'<l}(a_{l'}-1)i_1\right]}=A_{i_1,\ldots,i_p}$, with the convention that a product over the empty set equals $1$. 
By convention, when an array $B$ is defined as an $((a_1,\ldots,a_p),(b_1,\ldots,b_q))$-sized array (with two vector of indexes), 
$\vec{A}$ is the matrix $\left[\vec{A}_{i,j}\right]_{i\in\{1,\ldots,\prod_{l=1}^pa_l\},j\in\{1,\ldots,\prod_{l=1}^q b_l\}}$ 
such that $\forall (i_1,\ldots,i_p)\in \prod_{l=1}^p\{1,\ldots,a_l\}$, $(j_1,\ldots,j_q)\in \prod_{l=1}^p\{1,\ldots,a_l\}$,
\noindent $\vec{A}_{1+\sum_{l=1}^p \left[(i_l-1)\prod_{l'<l}(a_{l'})\right],1+\sum_{l=1}^q \left[(j_l-1)\prod_{l'<l}(b_{l'})\right]}=A_{(i_1,\ldots,i_p),(j_1,\ldots,j_p)}$. \newversion{Given $A$ an $((a_1,\ldots, a_n),(b_1,\ldots b_l))$ array and $B$ a $((b_1,\ldots, b_l),(c_1,\ldots c_p))$ array,
$C=A\times B$ is the $((a_1,\ldots, a_n),(c_1,\ldots c_p))$ array defined by $$C_{(i_1,\ldots,i_n),(k_1,\ldots,k_n)}=\sum_{j_1,\ldots, j_l}A_{(i_1,\ldots,i_n),(j_1,\ldots, j_l)}B_{(j_1,\ldots,j_l),(k_1,\ldots ,k_n)}.$$}

\subsection{The sample design}
The CPS monthly sample comprises about 72,000 housing
units and is collected for 729 areas (Primary Sampling Units) consisting of more than 1,000 counties
covering every state and the District of Columbia. The CPS, conducted by the Census Bureau, uses a 4-8-4 rotating
panel design. For any given
month, the CPS sample can be grouped into eight subsamples corresponding to the
eight rotation groups. All the units belonging to a particular rotating panel enter and
leave the sample at the same time. A given rotating panel (or group) stays in the sample for four
consecutive months, leaves the sample for the eight succeeding months, and
then returns for another four consecutive months. It is then dropped from the sample
completely and is replaced by a group of nearby households. Of the two new rotation
groups that are sampled each month, one is completely new (their first appearance
in the panel) and the other is a returning group, which has been out of the sample
for eight months. Thus, in the CPS design
six and four out of the eight rotation groups are common between two consecutive months (i.e.,
75\% overlap) and the same month of two consecutive years (i.e., 50\%
overlap) respectively; see \cite{Hansen1955}.
For  month $\mon$, let  $\sample_\mon$\index[notations]{Sm@$\sample_\mon$ : sample for month $\mon$} denote the sample of respondents.
Let $\sample_{\mon,\misi}$\index[notations]{Smg@$\sample_{\mon,\misi}$ : sample rotation group $\misi$ for month $\mon$} denote the set of sampled respondents in the $\misi$th
\index[notations]{g@$\misi$ : month-in-sample  index, $\misi=1,\ldots,8$} sample rotation group for month $\mon$
and  $\sample_\mon=\bigcup_{\misi=1}^8\sample_{\mon,\misi}$. For a given month $\mon$, the rotation groups $S_{\mon,\misi}$, $\misi=1,\ldots,8$ are indexed so that 
$\misi$ indicates the number of times that rotation group $\sample_{\mon,\misi}$ 
has been a part of the sample in month $\mon$ and before. In the US Census Bureau terminology, $\misi$ is referred to as the month-in-sample ($\mis$) index and $S_{\mon,\misi}$ as the month-in-sample $\misi$ rotation group (more details on the design are given in Section \ref{sec:4.3}). We adopt a design-based approach in this study in which variables $\paramx$ and $\paramy$ are considered fixed parameters of the underlying fixed population model for design-based inference \citep[p.~2]{CasselSarndalWretman1977}. 
\section{Estimation}

\subsection{Direct and month-in-sample  estimators}

Let $\paramw$ be the $(M,N)$-sized array indexed by $\mon$ and $\indiv$ where $\paramw_{\mon,\indiv}$ denotes the second stage weight of individual $\indiv$ in month $\mon$ 
(by convention, $\paramw_{\mon,\indiv}=0$ if $\indiv\notin \sample_\mon$),
\index[notations]{wmk@$\paramw_{\mon,\indiv}$ :  second-stage weight for the $\indiv$th individual in month $\mon$}
which is obtained from the basic weight (that is, the reciprocal of the inclusion probability) after standard non-response and post-stratification adjustments  (for more details, we refer to \cite{CPS2006}).
The array of direct survey-weighted estimator of $\total_\paramy $ is given by $\hat{\total}^{\direct}_\paramy =\left[\sum_{\indiv\in \sample_{\mon}}\paramw_{\mon,\indiv}\paramy_{\mon,\indiv,\status}\right]_{\mon\in\{1,\ldots,\Mon\},\status\in\{1,2,3\}}$
\index[notations]{t2@$\hat{\total}^\star_\paramy$ : a random $(\Mon, 3)$-sized array, estimator of $\total_\paramy$}.
\index[notations]{t2@$\hat{\total}^\star_\paramy$ : a random $(\Mon, 3)$-sized array, estimator of $\total_\paramy$!t3@$\hat{\total}^\direct_\paramy$ : direct estimator}
Define the $(\Mon,8,3)$-sized array of month-in-sample estimates: 
$\hat{\total}^{\mis}_\paramy =\left[8\times\sum_{\indiv\in \sample_{\mon,\misi}}\paramw_{\mon,\indiv}\paramy_{\mon,\indiv,\status}\right]_{\mon\in\{1,\ldots,\Mon\},\misi\in\{1,\ldots,8\},\status\in\{1,2,3\}}.$
\index[notations]{t2@$\hat{\total}^\star_\paramy$ : a random $(\Mon, 3)$-sized array, estimator of $\total_\paramy$!t4@$\left(\hat{\total}^\mis_\paramy\right)_{.,\misi,.}$, $\misi=1,\ldots,8$ : month-in-sample  $\misi$ estimator} For a month-in-sample number $\misi$, $\left(\hat{\total}^{\mis}_\paramy\right)_{.,\misi,.}$ is called the month-in-sample $\misi$ estimator of $\total_\paramy$.

\subsection{An extended Bailar model for the rotation group bias}
Because of differential non-response and measurement errors across different rotation groups, the direct and month-in-sample estimators are subject to a bias, commonly referred to as the rotation group bias.
\cite{Bailar1975} proposed a class of semi-parametric models on the expected values of the month-in-sample estimators. Under a model in this class, (i) the bias of each month-in-sample 
estimator of total of unemployed depends on the month-in-sample index $\misi$ only, (ii) the bias  is invariant with time, and (iii) the vector of month-in-sample biases are bounded by a known linear constraint (without this binding linear constraint, month-in-sample rotation group biases could only be estimated up to an additive constant). Note that these very strong assumptions were made in order to reveal the existence of what in the US Census Bureau terminology is known as the rotation group bias. It would be highly questionable 
to use this model for rotation group bias correction, because (i) the choice of the linear constraint  would be totally arbitrary  in the absence of a re-interview experiment and (ii) the stationarity assumptions are unreasonable. 
We propose the following model in order to extend the Bailar model  to  account for the rotation group biases of the multiple categories: 
\begin{equation}\mathrm{E}\left[\left(\hat{\total}_\paramy^{\mis}\right)_{\mon,\misi,\status}\right]=\left(\total_\paramy \right)_{\mon,\status}+b_{\misi,\status},\label{model:bailar}
\end{equation}
where $b$ is a two-dimensional  $(8,p)$-sized array of biases such that $\forall \status, C_\status b_{.,\status}=0$,  $C_1,C_2,C_3$ being known linear forms satisfying $C_\status\partransp{1,\ldots,1}\neq 0$.
\index[notations]{b@$b$ : $(8,3)$-sized matrix indexed by month-in-sample  and employment status, $b_{\misi,\status}$ is the bias of all month-in-sample  $\misi$ estimator of total of population with employment status $\status$ over the months in general Bailar model. vector of rotation group biases}

\subsection{Estimation of unemployment rate and variance approximation}\label{varur}
We define the function $\urf:(0,+\infty)^3\to [0,1], x\mapsto x_2/(x_1+x_2)$. By convention, when applied to an array with employment status as an index, $x_1$, $x_2$ denotes the subarrays for employment status 1 and 2, respectively, and 
$/$ denotes the term by term division. 
The unemployment rate vector is defined as $\ur=\urf\left(\total_\paramy\right)=\left(\total_\paramy\right)_{.,1}/\left(\left(\total_\paramy\right)_{.,1}+\left(\total_\paramy\right)_{.,2}\right)$.
\index[notations]{r@$\urf$ : function that returns unemployment rate from employment status frequencies}
\index[notations]{r2@$\ur$ : $\Mon$-sized vector indexed by month, $\ur_\mon$ is the unemployment rate for month $\mon$}
Given an estimator $\hat{\total}_\paramy^\star$\index[notations]{$\star$ : index of estimator type: direct , AK, $\rc$ (regression composite),$\mis$ (month-in-sample  )} of $\total_\paramy $,
we derive the following estimator of $\ur$ from $\hat{\total}_\paramy^\star$: $\hat{\ur}^\star=\urf(\hat{\total}_\paramy ^\star)$.\index[notations]{ur@$\hat{\ur}$ : $\Mon$-sized vector, estimator of unemployment rate derived from estimator of total of employed and unemployed, $\hat{\ur}^{\star}=\urf\left(\hat{\total}_\paramy^{\star}\right)$}
Using the linearization technique, we can approximate the variance $\mathrm{Var}\left[\hat{\ur}^\star_{\mon}\right]$ 
of the unemployment rate estimator for month $m$ by
$J_1\mathrm{Var}\left[\left(\hat{\total}^{\star}_\paramy \right)_{\mon,.}\right]\transp{J_1},$
where $J_1$ is the Jacobian matrix:
$J_1=\left(\frac{\mathrm{d}~\urf(t)}{\mathrm{d}~t}\right)\left((\total_\paramy )_{\mon,.}^\star\right)=\begin{bmatrix}\left(\total_\paramy \right)_{\mon,1}^{-1},-\left(\total_\paramy \right)_{\mon,1}\left(\total_\paramy \right)_{\mon,2}^{-2},0\end{bmatrix}$,
and the variance of the estimator of change of the employment rate between two consecutive months  by
$J_2\mathrm{Var}\left[\left(\left(\hat{\total}^{\star}_\paramy \right)_{\mon,.},\left(\hat{\total}^{\star}_\paramy \right)_{\mon-1,.}\right)\right]\transp{J_2},$
\index[notations]{J2@$J, J_1, J_2$ : Jacobian matrices}
where
\begin{eqnarray*}J_2&=&\left(\frac{\mathrm{d}~\urf(t)-\urf(t')}{\mathrm{d}~(t,t')}\left(\left(\total_\paramy \right)_{\mon,.},\left(\total_\paramy \right)_{\mon-1,.}\right)\right)\\
&=&\begin{bmatrix}\left(\total_\paramy \right)_{\mon,1}^{-1},-\left(\total_\paramy \right)_{\mon,1}\left(\total_\paramy \right)_{\mon,2}^{-2},0,-\left(\total_\paramy \right)_{\mon-1,1}^{-1},\left(\total_\paramy \right)_{\mon-1,1}\left(\left(\total_\paramy \right)_{\mon-1,2}\right)^{-2},0\end{bmatrix}.\end{eqnarray*}

\subsection{The class of linear combinations of month-in-sample  estimators}\label{sec:3.5}

Here, as in \cite{yansaneh1998optimal}, we consider the best estimator of counts by employment status in the class of linear combinations of month-in-sample  estimators.
Generalizing \cite{yansaneh1998optimal}, the unbiasedness assumption of all month-in-sample  estimators is:
\begin{equation}\mathrm{E}\left[{\vec{\hat{\total}}^{\mis}_\paramy}\right]=\vec{\Xmatrix}  {\vec\total_\paramy},\label{M1}\end{equation} where 
$X$\index[notations]{X@$X$ : a $((\Mon, 8, 3), (\Mon, 3))$-sized array} is  the $((\Mon, 8, 3), (\Mon, 3))$-sized array with rows indexed by the triplet $(\mon,\misi,\status)$ and columns indexed by the couple
$(\mon,\status)$ such that $X_{(\mon, \misi,\status),(\mon',\status')}=1$ if $\mon'=\mon$ and $\status=\status'$, $0$ otherwise.
Let  $L$\index[notations]{@,$L$ : a $(p,( \Mon, 3))$-sized array} be a $(p, (\Mon, 3))$-sized array  with $p\in\mathbb{N}\setminus \{0\}$ and rows indexed by $(\mon,\status)$. By class of linear 
estimators of $L\total_\paramy$, we will designate the class of estimators that are linear combinations of the month-in-sample  estimates, i.e. of the form $W \vec{\hat{\total}}^{\mis}_\paramy$ where
$W$ \index[notations]{W@$W$ : a $(( \Mon, 3),( \Mon, 8, 3))$-sized array of weights for a weighted combination of month-in-sample estimates} 
is a fixed (does not depend on the observations) $(p, (\Mon\times 8\times 3))$-sized matrix.

\subsubsection*{Best linear estimator}

Let $\Sigma_\paramy=\mathrm{Var}_\paramy\left[{\vec{\hat{\total}}^{\mis}_\paramy}\right]$. \index[notations]{Sigma@$\Sigma_\paramy$ : variance covariance  matrix, $\Sigma_\paramy=\mathrm{Var}_\paramy[\hat{\total}^{\mis}_\paramy]$} 
In the design-based approach, $\Sigma_\paramy$ is a function of the parameter $\paramy$.
The variance of a linear transformation $W  \vec{\hat{\total}}^{\mis}_\paramy$ of $\hat{\total}^{\mis}_\paramy$ is:
$\mathrm{Var}\left[W  \vec{\hat{\total}}^{\mis}_\paramy\right]=W^T  \Sigma_\paramy  W.$
When month-in-sample estimates are unbiased, $\Sigma_\paramy$ is known, and only $\vec{\hat{\total}}^{\mis}_\paramy$ is observed, and \newversion{when $\vec\Xmatrix^+\vec\Xmatrix=I$}, the Gauss-Markov theorem states that 
the BLUE of $\total_\paramy$ uniformly in $\total_y$ is the $(\Mon,3)$-sized matrix $\hat{\total}^{\text{BLUE}}_\paramy$ defined by

\newversion{
\begin{equation}\label{bestW}
\vec\Xmatrix ^+ (\vec\Xmatrix   \vec\Xmatrix ^+)  \left(I-\Sigma_\paramy ((I-\vec\Xmatrix   \vec\Xmatrix ^+)^+ \Sigma_\paramy  (I-\vec\Xmatrix   \vec\Xmatrix ^+))^+\right)\vec{\hat{\total}}^{\mis}_\paramy,
\end{equation}
}

\noindent
where the $^+$\index[notations]{=@${}^+$ : operator, Moore Penrose pseudo inverse} operator designates the Moore Penrose pseudo inversion, $I$ is the
identity matrix. Here the minimisation is with respect to the order on symmetric positive definite matrices: $M_1\leq M_2 \Leftrightarrow M_2-M_1$ is positive. It can be shown that $\vec\Xmatrix ^+=\transp{\vec\Xmatrix }/8$ in our case and that $\vec\Xmatrix ^+\vec\Xmatrix=I$. For more details about the Gauss-Markov result under singular linear model, one may refer to \newversion{\cite[p.~140, Eq. 3b]{Searle1994}}.
This is a generalization of the  result  of \cite{yansaneh1998optimal}, as it takes into account the multi-dimensions of $\paramy$ and non-invertibility of $\Sigma_\paramy$.
Note that $\Sigma_\paramy$ can be non-invertible, especially when the sample is calibrated on a given fixed population size, considered non-random, because of an affine relationship between month-in-sample  estimates (e.g., $\sum_{\misi=1}^8\sum_{\status=1}^3 \left(\hat{\total}^{\mis}_\paramy \right)_{\mon,\misi,\status} $ is not random). 

It is important to recall that (i) for any linear transformation $L$ applicable to $\vec{\total}_\paramy$, the best linear unbiased estimator of $L {\vec{\total}_\paramy}$  uniformly in $\total_\paramy$ is $L  \vec{\hat{\total}}_\paramy^{\text{BLUE}}$, which ensures that the BLUE of month-to-month change can be simply obtained from the BLUE of level and so there is no need for searching a compromise between estimation of level and  change;
(ii) for any linear transformation $L$ applicable to $\vec{\total}_\paramy$, any linear transformation $J$ applicable to $L\vec{\total}_\paramy$,  $L\vec{\hat{\total}}_\paramy^{\text{BLUE}}\in \argmin\left\{\left.J W \Sigma_\paramy \partransp{J W }\right|W, W\vec\Xmatrix =L\right\}$, which ensures that plug-in estimators for unemployment rate and month-to-month unemployment rate change derived from the BLUE are also optimal 
in the sense that they minimize the linearized approximation of the variance of such plug-in estimators, that can be written in the form $J W \Sigma_\paramy \partransp{J W }$.

\subsubsection*{Remark: BLUE under Bailar rotation bias model}
Here we give the expression of the BLUE under the general Bailar rotation bias model.

Bailar's rotation bias model can be written in matrix notation:
\begin{equation}\mathrm{E}\left[\vec{\hat{\total}}^{\mis}_\paramy\right]=\vec\Xmatrix   \vec\total_\paramy+ \vec\Xmatrix'  \vec{\rotationbias},\label{bailarmatrix}
 \end{equation}
\index[notations]{X'@$\Xmatrix'$ : a matrix} where $\Xmatrix'$ is a fixed known array (see also
\cite[equation 8]{yansaneh1998optimal}).
For example under Model \eqref{model:bailar}, with $C_1=C_2=C_3=(1,\ldots,1)$, 
$\Xmatrix'$ is the  $((\Mon, 8, 3), (7, 2))$-sized array where
for $\mon\in\{1,\ldots,\Mon\}$, $\misi \in \{1,\ldots,8\}$, $\misi' \in \{1,\ldots,7\},\status\in\{1,2,3\}  $, $\status'\in\{1,2,3\}$,
$\Xmatrix'_{(\mon,\misi,\status),(\misi',\status')}=
1$  if $\misi=\misi'<8$  and  $\status=\status'$, $-1$  if $\misi=8$  and $\status=\status',$ $0$ otherwise.
We can reparametrize Model \eqref{bailarmatrix} in the form $\mathrm{E}[\hat{\total}^{\mis}_\paramy]=X^\star  \mu$, where $X^\star=[\vec\Xmatrix \mid \vec\Xmatrix ']$, and the parameter $\mu=\transp{[\vec\total_\paramy\mid  \vec{\rotationbias}]}$. The best linear unbiased  estimator of $\vec{\total}_\paramy$ under this rotation bias model is

\newversion{
$$L X^{\star+}(X^\star X^{\star+})\left(I-\Sigma_\paramy(I-X^\star X^{\star+})^+\Sigma_\paramy (I-X^\star X^{\star+}))^+\right)\vec{\hat{\total}}^{\mis}_\paramy,$$}
 with  $L$ satisfying $LX^\star=\vec\Xmatrix $. 
This is a generalization of \cite{yansaneh1998optimal}, as it considers non-invertible $\Sigma_\paramy$, does not limit to a unidimensional variable and is generalized to general Bailar's model.

\subsection{AK composite estimation}
\subsubsection*{Definition}

We define a general class of  AK composite estimators. Let  $A=\mathrm{diag}(a_1,a_2,a_3), K=\mathrm{diag}(\indiv_1,\indiv_2,\indiv_3)$, be real diagonal $(3, 3)$ matrices, 
 The AK estimator with coefficients $A$ and $K$ is defined as follows: first define
$\left(\hat{\total}^{\ak}_\paramy\right)_{1,.}=\left(\hat{\total}^{\direct}_\paramy\right)_{1,.},$
then recursively define for $\mon\in 2,\ldots,\Mon$,  
\begin{multline}
\left(\hat{\total}^{\ak}_\paramy\right)_{\mon,.}=K\left(\hat{\total}^{\direct}_\paramy\right)_{\mon,.}\\
+ (I-K)\times\left(\left( \hat{\total}^{\ak}_\paramy\right)_{\mon-1,.}+\sum_{\indiv\in \sample_\mon\cap \sample_{\mon-1}}\left(\paramw_{\mon,\indiv,.}\paramy_{\mon,\indiv,.}-\paramw_{\mon-1,\indiv,.}\paramy_{\mon-1,\indiv,.}\right)\right)\\
+A\times\left(\sum_{\indiv\in \sample_\mon\setminus \sample_{\mon-1}}\paramw_{\mon,\indiv,.}\paramy_{\mon,\indiv,.}-\frac13\sum_{\indiv\in \sample_\mon\cap \sample_{\mon-1}}\paramw_{\mon,\indiv,.}\paramy_{\mon,\indiv,.} \right),
\end{multline}
\index[notations]{t2@$\hat{\total}^\star_\paramy$ : a random $(\Mon, 3)$-sized array, estimator of $\total_\paramy$!t5@$\hat{\total}^\ak_\paramy$ : AK estimator}
where $\setminus$ denotes the set difference operator and $I$ is the identity matrix of dimension 3.
The sum of the first two terms of the AK estimator is indeed  a weighted average of the current month direct estimator and  the previous month AK estimator suitably updated for the  change. The last term of the AK estimator is correlated to the previous terms, and has an expectation 0 with respect to the sample design. \cite{gurney1965multivariate} explained the benefits of adding the third term in reducing the  mean squared error. The Census Bureau uses specific values of $A$ and $K$, which were empirically determined in order to arrive at a compromise solution that worked reasonably well for both employment level and rate estimation (see. \cite{lent1999effect}).
The corresponding  unemployment rate estimator is obtained as:
$\hat{\ur}_\mon^{\ak}=\urf\left(\left(\hat{\total}^{\ak}_\paramy\right)_{\mon,.}\right)$.
Note that $\hat{\ur}_\mon^{\ak}$ just depends on $a_1$, $a_2$, $k_1$, $k_2$ and not on
$a_3$ and $k_3$.
Note that the class of AK estimators is a sub class of the class of linear estimators, as the AK estimator can be written as a linear  combination of the month-in-sample  estimators: 
$\left(\hat{\total}^{\ak}_\paramy \right)_{\mon,.}=\sum_{\mon'=0}^\mon\sum_{\misi =1}^8 c_{\mon,\mon',\misi}\left(\hat{\total}^{\mis}_\paramy\right)_{\mon',\misi,.},$
where the $(3,3)$ matrices  $c_{\mon,\mon,\misi}$ are defined recursively:
$\forall \misi \in \{1,\ldots,8\}, c_{0,0,\misi}=(1/8)\times\mathrm{I}_3,$ where $\mathrm{I}_3,$ is the $(3,3)$ identity matrix and 
\begin{equation}\label{cond2}
\forall \mon \in \{2,\ldots,\Mon\},
\left\{\begin{array}{lll}
     \forall \misi \in \{1,5\}& c_{\mon,\mon,\misi} &=( I -K)/8+A\\
     \forall \misi \in  \{2,3,4,6,7,8\}& c_{\mon,\mon,\misi} &=( I -K)/8+K/6-A/3\\
     \forall \misi \in \{1,2,3,5,6,7\}& c_{\mon,\mon-1,\misi} &=c_{\mon-1,\mon-1,\misi}\times K-K/6\\
     \forall \misi \in \{4,8\}& c_{\mon,\mon-1,\misi} &=c_{\mon-1,\mon-1,\misi}\times K\\
     \forall 1\leq\mon'<\mon-1& c_{\mon,\mon',\misi} &=c_{\mon-1,\mon',\misi}\times K\\
  \end{array}\right.
\end{equation}
$\forall \mon'>\mon,\misi \in \{1,\ldots,8\}, c_{\mon,\mon',\misi}=0.$

Let  $W^{\ak}$ be the $((\Mon, 3),(\Mon, 8, 3))$ array, such that
for $\mon,\mon'\in \{1,\ldots,M\}$, $\misi\in\{1,\ldots,8\}$, $\status,\status'\in\{1,2,3\}$,
$W^{\ak}_{(\mon,\status),(\mon',\misi,\status')}=c_{\mon,\mon',\misi}$ if $\status=\status'$, $0$ otherwise. Then
$\vec{\hat{\total}}^{\ak}_\paramy =\vec{W}^{\ak} \vec{\hat{\total}}^{\mis}_\paramy$. \index[notations]{!w2@$W^{AK}$ : $((\Mon, 3),(\Mon, 8, 3))$-sized array of weights for the AK estimator}

\subsubsection*{Notes on AK estimator}

In presence of rotation bias, the bias of the AK estimator is not null and  equal to $\vec{W}^{\ak}\vec\Xmatrix '\vec{b}$. Depending on the rotation bias model, there may not exist an unbiased version of the AK estimator.
Furthermore, contrary to the BLUE, the best A,K coefficients for estimation of one particular month and status may not be optimal for another month and status, and the best A,K, coefficients for estimation of level may not be optimal for estimation of change.
For example, one may find $A,K,\mon,\status, A',K',\mon',\status'$ such that
$\mathrm{Var}\left[\left(\hat{\total}^{\ak}_\paramy\right)_{\mon,\status}\right]<\mathrm{Var}\left[\left(\hat{\total}^{\mathrm{A'},\mathrm{K'}}_\paramy\right)_{\mon,\status}\right]$ and
$\mathrm{Var}\left[\hat{\total}^{\ak}_{\paramy_{\mon',\status'}}\right]>\mathrm{Var}\left[\hat{\total}^{\mathrm{A'},\mathrm{K'}}_{\paramy_{\mon',\status'}}\right]$.
When $\Sigma_\paramy$ is known, let 
$\hat{\total}_{\paramy}^{BAK,level}$,
$\hat{\total}_{\paramy}^{BAK,change}$, 
$\hat{\total}_{\paramy}^{BAK,compromise}$ be the AK estimators obtained for $A$, $K$, that minimize the average approximated variance of 
level estimates 
$\sum_{\mon=1}^\Mon J_1\mathrm{Var}_\paramy\left[\left(\hat{\total}^{A,K}_\paramy\right)_{\mon,.}\right]\transp{J_1}$, of change estimates
$\sum_{\mon=1}^\Mon J_2\mathrm{Var}_\paramy\left[\left(\hat{\total}^{A,K}_\paramy\right)_{\{\mon-1,\mon\},.}\right]\transp{J_2}$ and compromise averaged variance
$$\sum_{\mon=1}^\Mon \left(J_1\mathrm{Var}_\paramy\left[\left(\hat{\total}^{A,K}_\paramy\right)_{\mon,.}\right]\transp{J_1}+J_2\mathrm{Var}_\paramy\left[\left(\hat{\total}^{A,K}_\paramy\right)_{\{\mon-1,\mon\},.}\right]\transp{J_2}\right),$$ respectively.
For AK estimation, note that the three objective functions are polynomial functions of  $A$ and $K$ whose coefficients are functions of $\Sigma_\paramy$, and by using a standard numerical method (Nelder Mead) we can obtain the optimal coefficients. 

\subsection{Empirical best linear estimator and empirical best AK estimator.}

Let $\hat{\Sigma}$ be an estimator of $\Sigma_\paramy$, and 
let $\hat{\total}^{EBLUE}_\paramy$ be the estimator of $\total_\paramy$ obtained from \eqref{bestW} when $\Sigma_\paramy$ is replaced by $\hat{\Sigma}$. In the same manner, we can define the empirical best AK estimators for change, level and compromise.
For the CPS, optimal $A$ and $K$ coefficients were determined so that a compromise objective function, accounting for the variances  of the month-to-month changes and levels estimates, would be minimum. The variances were estimated according to a stationary covariance of month-in-sample  estimates assumption (see \cite{Lent1996}) and the method used in the Census Bureau consists in choosing the best coefficients $a_1$, $a_2$, $k_1$, $k_2$ on a grid with 10 possible values for each coefficient $(0.1,\ldots,0.9)$.

\subsection{Regression Composite Estimation}\label{RC}

In this section we elaborate on the general definition of the  class of regression composite estimators  proposed by \cite{fuller2001regression}, parametrized by a real number $\alpha\in[0,1]$. This class includes regression composite estimators MR1 (for $\alpha=0$) and MR2 (for $\alpha=1$) as defined by \cite{singh1995composite} and \cite{singh2001regression}.
For  $\alpha\in[0,1]$, the  regression composite estimator of $\total_{\paramy}$  is a calibration  estimator $\left(\hat{\total}^{\rcind,\alpha}_\paramy\right)_{\mon,.}$ defined as follows:
provide calibration totals $\left(\total^{adj}_{\paramx}\right)_{\mon,.}$ for the auxiliary variables (they can be equal to the true totals when known or estimated), then 
define $ \left(\hat{\total}^{\rcind ,\alpha}_\paramz\right)_{1,.}=\left(\hat{\total}^{\direct}_\paramz\right)_{1,.},$  and  $\paramw_{1,\indiv}^{\rcind ,\alpha}=\paramw_{1,\indiv}$ if $k\in \sample_1$, 0 otherwise.
For $\mon \in \{2,\ldots, \Mon\}$,  recursively define \index[notations]{MR@{MR1, MR2, MR3} : indicates the modified regression 1, 2 and 3 estimators}\index[notations]{alpha@$\alpha$ : coefficient in $[0,1]$ used to defined the Fuller and Rao regression composite estimator}
\begin{align}
\paramzc[(\alpha)]_{\mon,\indiv,.}&=
  \begin{cases}
     \alpha\left(\tau_\mon^{-1}
          \left(\paramz_{\mon-1,\indiv,.}-\paramz_{\mon,\indiv,.}\right) +\paramz_{\mon,\indiv,.}\right)
     +(1-\alpha)~\paramz_{\mon-1,\indiv,.} & \text{if }k\in \sample_{\mon}\cap \sample_{\mon-1},
\\
 \alpha~ \paramz_{\mon,\indiv,.}
 +(1-\alpha)~\left(\sum_{k\in \sample_{\mon-1}}\paramw_{\mon-1,\indiv}^{\rcind ,\alpha}\right)^{-1}
\left(\hat{\total}_\paramy ^{\mathrm{c}}\right)_{\mon-1,.} & \text{if }k\in \sample_{\mon}\setminus \sample_{\mon-1},
\end{cases}\label{RCstep1}
\end{align}\index[notations]{zstar@$\paramzc$ : 3-dimensional $(\Mon,8,3)$-sized array of proxy variables for $\paramz$ defined for the regression composite estimator.}
where
$\tau_\mon=\left(\sum_{k\in \sample_\mon\cap \sample_{\mon-1}}\paramw_{\mon,\indiv}\right)^{-1}\sum_{k\in \sample_\mon}\paramw_{\mon,\indiv}$.
Then the regression composite estimator of $\left(\total_{\paramy}\right)_{\mon,.}$ is given by
$\left(\hat{\total}^{\rcind,\alpha}_\paramy\right)_{\mon,.}=
\sum_{k\in \sample_\mon}\paramw^{\rcind,\alpha}_{\mon,\indiv}\paramy_{\mon,\indiv},$
where
\begin{equation}
\left(\paramw^{\rcind,\alpha}_{\mon,.}\right)\!=\!\argmin\left\{\sum_{k\in U}\frac{ \left(\paramw^\star_{k}-\paramw_{\mon,\indiv}\right)^2}{\mathds{1}(k\notin S_m)+\paramw_{\mon,\indiv}}\left|
\paramw^\star\in\mathbb{R}^{U},\!\!\!
\begin{array}{l}\sum_{k\in \sample_\mon} \paramw^\star_{k}\paramzc[(\alpha)]_{\mon,\indiv,.}\!=\!\left(\hat{\total}^{\rcind,\alpha}_\paramz\right)_{\mon-1,.}\\
\sum_{k\in \sample_\mon} \paramw^\star_{k}\paramx_{\mon,\indiv,.}=\left(\total^{adj}_{\paramx}\right)_{\mon,.}
\end{array}
\right.\!\!\!\! \right\}\!\!,\label{RCstep2}\end{equation}
and
$\left(\hat{\total}^{\rcind,\alpha}_\paramz\right)_{\mon,.}=
\sum_{k\in \sample_\mon}\paramw^{\rcind,\alpha}_{\mon,\indiv}\paramzc[(\alpha)]_{\mon,\indiv},$
where $\mathds{1}(k\notin S_m)=1$ if $k\notin S_m$ and $0$ otherwise.
Our definition of regression composite estimator is more general than in \cite{fuller2001regression} as it takes into account a multivariate version of $\paramy$.
Modified Regression 3 (MR3), of \cite{gambino2001regression},  does not belong to the class of regression composite estimators. The MR3 estimator imposes too many constraints in the calibration procedure, which leads to a high variability of the calibration weights, and consequently, MR3 estimator has a larger MSE than composite regression estimators.

\subsubsection*{Choice of $z$ and  choice of $\alpha$}
\cite{fuller2001regression} studied the properties of the estimator
$\left(\hat{\total}^{\rcind,\alpha}_\paramy\right)_{\mon,1}$ for the  choice of
$\paramz=\paramy_{.,1}$.
As the employment rate is a function of $\paramy_{\mon,1}$ and $\paramy_{\mon,2}$,
we studied the properties of Regression Composite Estimator with the choice
  $\paramz=\paramy$.
 \cite{fuller2001regression} proposed a method that allows an approximation of the optimal $\alpha$ coefficient for  month-to-month change and level estimation, under a specific  individual level superpopulation model for continuous variables.
They proposed this superpopulation model to  explain the drift problem of MR2 (regression composite estimator for  $\alpha=1$) and obtain the  best coefficient $\alpha$. Since we deal with a discrete multidimensional variable, the continuous superpopulation model assumed by \cite{fuller2001regression} is not appropriate in our situation. It will be interesting to propose an approach to estimate the best $\alpha$ in our situation. For our preliminary study we examined a range of known $\alpha$ values in our simulations and in the CPS data analysis.

\section{Simulation Experiment}

\subsection{Description of Simulation Study}
We conducted a simulation study  to enhance our understanding of the finite sample properties of different composite estimators.   We generated  three finite populations, each with size 100,000.
In order to make the simulation experiment meaningful, we generated employment statuses for each finite population in a manner that  attempts to capture the actual U.S. national employment rate dynamics during the study period 2005-2012. Moreover, in order to understand the maximum gain from the composite estimation, we induced high correlation in the employment statuses between two consecutive months subject to a constraint on the global employment rate evolution.  We set the probability of month-to-month changes in employment statuses for an
individual to zero in case of no change in the corresponding direct  national employment rates.
Samples where selected according to a rotating design with systematic selection that mimics the CPS design. Since the number of possible samples is only 1000, we are able to compute the exact design-based  bias,  variance and mean squared error  of  different estimators, and, subsequently, the optimal linear and optimal AK estimators.
We compute employment rate, total employed, and total unemployed  series over the 85-month period using the direct, AK and the Fuller-Rao  regression composite methods.  We then compared the optimal estimator in the class of  regression composite estimators to those in the class of the AK  and best linear estimators.
{Note that the simulation study can be reproduced using the R package we created for this purpose (see \cite{github:pubBonneryChengLahiri2016}).}
\subsection{Populations generation}
We created three populations of $N=100,000$ individuals each, indexed by $1,\ldots,N$.
For each individual $k$ of each population, we created a time series $(\paramy_{\mon,\indiv,.})_{\mon\in{1,\ldots,\Mon}}$,
where $\paramy_{\mon,\indiv,.} \in \{(1,0,0),(0,1,0),(0,0,1)\}$ (for unemployed, not in labor force, employed), and with $M=85$\index[notations]{M@$\Mon$ : total number of months, equal to $85$ in the simulations and in the CPS data study}.
Each individual belongs to one household. Each household consists
of $5$ individuals. The number of all households is $\numofhinpop=20,000$, the set of all households is $\left\{\household_i=\left\{(5\times(i-1)+1),\ldots, (5\times i) \right\}\mid i=1,\ldots,\numofhinpop\right\}$
\index[notations]{i@$i$ : index of the households in the simulations}\index[notations]{H@$H$ : number of households in the simulations ($H=20.000$)}  \index[notations]{hi@$h_i$ : household $i$}.
The time series are created under certain constraints at the population level.
For each population, the unemployment rates are the same as the direct estimates obtained from the CPS data.
In population 1, the number of people who change status between two consecutive months is minimal.
In populations 2 and 3, the proportions of persons who change from one status to another between two consecutive months are equal to those proportions as estimated from the CPS data.
In population 2, people with a small index have a higher probability to change status, whereas the probability to change status between two months is the same for all  individuals of population 3 with a same status.

\subsection{Repeated design}\label{sec:4.3}

We mimic the CPS design, which is described in appendix \ref{ap:cps}. For  month $\mon$, a sample $\sample_{\mon}$ is the union of 8 rotation
groups. {The design and the creation of} rotation groups are explained below.
Rotation groups are made of $\numofhinrg=20$ households, i.e. 100 individuals.
So for each month $\mon$, there are $\#(\sample_{\mon})=800$ individuals in the sample, and the inclusion probability of any unit is
$1/125$.
{The selection of longitudinal sample $S_1,\ldots S_m$ is made in 3 steps}:
\begin{enumerate}
\item Draw an integer number $\random$ between 1 and 1,000, from a uniform
distribution.
\item For $\ell\in{1,\ldots,(\Mon+15)},$  create
the cluster of households
$\rotationgroup_{\ell}=\bigcup_{j =1}^{\numofhinrg}\household_{i_{\ell,j}}$,\index[notations]{clusterl@$\rotationgroup_\ell$ : cluster of households $\ell$}\index[notations]{l@$\ell$ : cluster index}
\index[notations]{h@$\numofhinrg$ : number of households in a rotation group in the simulations ($\numofhinrg=20$)}
 where $i_{\ell,j}=\mathrm{rem}\left((r-1+\ell-1)+\frac{\numofhinpop}{\numofhinrg}\times(j-1),\numofhinpop\right)+1$,
and $\mathrm{rem}(a,b)$ denotes the remainder of the Euclidean division of $a$ by $b$.
\item Let $\delta_1=0$,$\delta_2=1$, $\delta_3=2$, $\delta_4=3$, $\delta_5=12$, $\delta_6=13$, $\delta_7=14$, $\delta_8=15$. For $\mon\in \left\{1,\ldots,\Mon\right\}$, $\misi \in \{1,\ldots,8\}$,  create the samples
$\sample_{\mon,\misi}=\rotationgroup_{\mon+\delta_\misi},$\index[notations]{delta@$\delta=(\delta_1,\ldots,\delta_8)$ : a vector for CPS rotation group lag : $\delta_6=13$ means that $13$
months after being rotation group $1$,  a cluster is rotation group $6$, by the relation $\sample_{\mon,\misi}=\rotationgroup_{\mon+\delta_g}$}
and $\sample_{\mon}=\bigcup_{\misi=1}^8S_{\mon,\misi}.$
\end{enumerate}

{As only 1000 different possible samples exist, we will be able in our simulation to draw them all and to compute exact design-based moments.
Table \ref{tab:cpsrotchar} displays the rotation chart for our simulation, which is identical to the CPS rotation chart \cite[Figure 3-1]{CPS2006}.
\setlength\tabcolsep{.2pt}
\begin{table}[h] \scriptsize \centering          \caption{Rotation chart}\label{tab:cpsrotchar}
\begin{tabular}{@{\extracolsep{0pt}} ccccccccccccccccccccc}  \\
\hline  &$\rotationgroup_{1}$&$\rotationgroup_{2}$&$\rotationgroup_{3}$&$\rotationgroup_{4}$&$\rotationgroup_{5}$&$\rotationgroup_{6}$&$\rotationgroup_{7}$&$\rotationgroup_{8}$&$\rotationgroup_{9}$&$\rotationgroup_{10}$&$\rotationgroup_{11}$&$\rotationgroup_{12}$&$\rotationgroup_{13}$&$\rotationgroup_{14}$&$\rotationgroup_{15}$&$\rotationgroup_{16}$&$\rotationgroup_{17}$&$\rotationgroup_{18}$&$\rotationgroup_{19}$&$\rotationgroup_{20}$\\
\hline \\Jan 05 & $S_{1,1}$ & $S_{1,2}$ & $S_{1,3}$ & $S_{1,4}$ &  &  &  &  &  &  &  &  & $S_{1,5}$ & $S_{1,6}$ & $S_{1,7}$ & $S_{1,8}$ &  &  &  &  \\  Feb 05 &  & $S_{2,1}$ & $S_{2,2}$ & $S_{2,3}$ & $S_{2,4}$ &  &  &  &  &  &  &  &  & $S_{2,5}$ & $S_{2,6}$ & $S_{2,7}$ & $S_{2,8}$ &  &  &  \\  Mar 05 &  &  & $S_{3,1}$ & $S_{3,2}$ & $S_{3,3}$ & $S_{3,4}$ &  &  &  &  &  &  &  &  & $S_{3,5}$ & $S_{3,6}$ & $S_{3,7}$ & $S_{3,8}$ &  &  \\  Apr 05 &  &  &  & $S_{4,1}$ & $S_{4,2}$ & $S_{4,3}$ & $S_{4,4}$ &  &  &  &  &  &  &  &  & $S_{4,5}$ & $S_{4,6}$ & $S_{4,7}$ & $S_{4,8}$ &  \\  May 05 &  &  &  &  & $S_{5,1}$ & $S_{5,2}$ & $S_{5,3}$ & $S_{5,4}$ &  &  &  &  &  &  &  &  & $S_{5,5}$ & $S_{5,6}$ & $S_{5,7}$ & $S_{5,8}$ \\  Jun 05 &  &  &  &  &  & $S_{6,1}$ & $S_{6,2}$ & $S_{6,3}$ & $S_{6,4}$ &  &  &  &  &  &  &  &  & $S_{6,5}$ & $S_{6,6}$ & $S_{6,7}$ \\  Jul 05 &  &  &  &  &  &  & $S_{7,1}$ & $S_{7,2}$ & $S_{7,3}$ & $S_{7,4}$ &  &  &  &  &  &  &  &  & $S_{7,5}$ & $S_{7,6}$ \\  Aug 05 &  &  &  &  &  &  &  & $S_{8,1}$ & $S_{8,2}$ & $S_{8,3}$ & $S_{8,4}$ &  &  &  &  &  &  &  &  & $S_{8,5}$ \\  Sep 05 &  &  &  &  &  &  &  &  & $S_{9,1}$ & $S_{9,2}$ & $S_{9,3}$ & $S_{9,4}$ &  &  &  &  &  &  &  &  \\  Oct 05 &  &  &  &  &  &  &  &  &  & $S_{10,1}$ & $S_{10,2}$ & $S_{10,3}$ & $S_{10,4}$ &  &  &  &  &  &  &  \\  Nov 05 &  &  &  &  &  &  &  &  &  &  & $S_{11,1}$ & $S_{11,2}$ & $S_{11,3}$ & $S_{11,4}$ &  &  &  &  &  &  \\  Dec 05 &  &  &  &  &  &  &  &  &  &  &  & $S_{12,1}$ & $S_{12,2}$ & $S_{12,3}$ & $S_{12,4}$ &  &  &  &  &  \\  Jan 06 &  &  &  &  &  &  &  &  &  &  &  &  & $S_{13,1}$ & $S_{13,2}$ & $S_{13,3}$ & $S_{13,4}$ &  &  &  &  \\  Feb 06 &  &  &  &  &  &  &  &  &  &  &  &  &  & $S_{14,1}$ & $S_{14,2}$ & $S_{14,3}$ & $S_{14,4}$ &  &  &  \\  Mar 06 &  &  &  &  &  &  &  &  &  &  &  &  &  &  & $S_{15,1}$ & $S_{15,2}$ & $S_{15,3}$ & $S_{15,4}$ &  &  \\  Apr 06 &  &  &  &  &  &  &  &  &  &  &  &  &  &  &  & $S_{16,1}$ & $S_{16,2}$ & $S_{16,3}$ & $S_{16,4}$ &  \\  May 06 &  &  &  &  &  &  &  &  &  &  &  &  &  &  &  &  & $S_{17,1}$ & $S_{17,2}$ & $S_{17,3}$ & $S_{17,4}$ \\  Jun 06 &  &  &  &  &  &  &  &  &  &  &  &  &  &  &  &  &  & $S_{18,1}$ & $S_{18,2}$ & $S_{18,3}$ \\  Jul 06 &  &  &  &  &  &  &  &  &  &  &  &  &  &  &  &  &  &  & $S_{19,1}$ & $S_{19,2}$ \\  Aug 06 &  &  &  &  &  &  &  &  &  &  &  &  &  &  &  &  &  &  &  & $S_{20,1}$ \\  \hline\end{tabular}  \end{table}

}
{
For example, for $\random=506$, $m=12$, $g=3$, we have $S_{m,g}=\rotationgroup_{12+\delta_3}=\rotationgroup_{14}$, and
$\rotationgroup_{14}=\{h_{\mathrm{rem}((506-1+14-1)+\frac{20000}{20}\times(k-1),20000)+1}\mid k=1\ldots 20\}=
\{h_{19},h_{1019},h_{2019},h_{3019},\ldots,h_{19019}\}$.
}

\subsection{Rotation bias}
In each sample, we introduced a measurement error  by changing employment status of $20\%$ of employed individuals in month-in-sample  group 1 from employed to unemployed,  which leads to an overestimation of  the unemployment rate.

\subsection{Variance on month-in-sample  estimators computation}
As we draw all the possible samples, we are able to compute  the exact variance of any estimator. Moreover, we are able to compute the true $\Sigma_\paramy$, which yields both the optimal best linear and AK estimators.

\subsection{Estimation of $\Sigma_\paramy$}
Define $$\sigma^2_{\mon,\mon'}=
\frac{\sum_{i=1}^ \numofhinpop
     \left(\sum_{\indiv\in h_i}\paramy_{\mon ,\indiv,.}-\frac{\sum_{i'=1}^ \numofhinpop\sum_{\indiv'\in h_{i'}}\paramy_{\mon,\indiv',.}}{\numofhinpop}\right)
\partransp{\sum_{\indiv\in h_i}\paramy_{\mon',\indiv,.}}}{\numofhinpop-1}.$$
We estimate $\sigma^2_{\mon,\mon'}$
 by $$\hat{\sigma}^2_{\mon,\mon'}=\frac{
            \sum\limits_{i\in\{1,\ldots,\numofhinpop\} \mid h_i\subset \sample_\mon\cap\sample_{\mon'}}
\left(\sum\limits_{\indiv\in h}\paramy_{\mon,\indiv,.}-\frac{\sum_{i=1}^ \numofhinpop\sum_{\indiv\in h_i}\paramy_{\mon',\indiv,.}}{\# \{i\in\{1,\ldots,\numofhinpop\} \mid h_i\subset \sample_\mon\cap\sample_{\mon'}\}}\right)\partransp{\sum\limits_{\indiv\in h}\paramy_{\mon',\indiv,.}}}{\#\left\{i \in \{1,\ldots,\numofhinpop\}\mid h_i\subset \sample_\mon\cap\sample_{\mon'}\right\}-1}$$ if $\sample_\mon\cap\sample_{\mon'}\neq\emptyset$, $0$ otherwise.
Let $\mon, \mon'\in \left\{1,\ldots,\Mon\right\}$,   $\misi,\misi'\in\{1,\ldots,8\}$.
If $m'+\delta_{\misi'}=m+\delta_{\misi}$ then
$S_{\mon,\misi}=S_{\mon',\misi'}$, we approximate the distribution of $S_{\mon',\misi'}$ by a cluster sampling, where first stage is simple random sampling.
and
we estimate
$\mathrm{Cov}\left[\hat{\total}^{\mis,\misi}_{\mon},\hat{\total}^{\mis,\misi}_{\mon'}\right]$
by $\widehat{\mathrm{Cov}}\left[\hat{\total}^{\mis,\misi}_{\paramy_{\mon,\status}},\hat{\total}^{\mis,\misi'}_{\paramy_{\mon',\status'}}\right]=(\numofhinpop)^2\left(1-\frac{\numofhinrg}{\numofhinpop}\right)\frac{\hat{\sigma}^2_{\mon,\mon'}}{\numofhinrg/8}.$
If $m'+\delta_{\misi'}\neq m+\delta_{\misi}$ then $\sample_{\mon,\misi}\cap\sample_{\mon',\misi'}=\emptyset$ and we approximate the distribution of  $(\sample_{\mon,\misi},\sample_{\mon',\misi'})$ by the distribution of two independent simple random samples of clusters conditional to non-overlap of the two samples, and we estimate
$\mathrm{Cov}\left[\hat{t}^{\mis}_{\mon,\misi,.},\hat{t}^{\mis}_{\mon',\misi',.}\right]$
by $\widehat{\mathrm{Cov}}\left[\hat{\total}^{\mis}_{\paramy_{\mon,\misi,.}},\hat{\total}^{\mis}_{\paramy_{\mon',\misi',.}}\right]=-\numofhinpop\hat{\sigma}^2_{\mon,\mon'}$.

\subsection{Choice of optimal estimator in each class}
In our simulations, the best linear unbiased estimator turned out to be exact, in the sense that for the three different choices of $\paramy$ (population 1, population 2, population 3), 
the $(1000,2040)$-matrix $Y$ whose rows are the $1000$ probable values of $\vec{\hat{\total}}^{\mis}_\paramy$ is of rank $1000$, so for all $(\mon,\status)$,  we can find a $2040$-sized vector  $x_{\mon,\status}$ such that $Yx_{\mon,\status}=\left(\total_\paramy\right)_{\mon,\status}.\mathds{1}$, where $\mathds{1}$ is $1000$-sized vector of ones. 
Then we define $W_o$ as the $((\Mon\times 8\times 3),(\Mon\times 3))$-sized array whose rows are the vectors $x_{\mon,\misi}$ such that $W_0\transp{Y}=\vec{\total}_\paramy$, which means that surely
$W_o\vec{\hat{\total}}^{\mis}_\paramy=\vec{\total}_\paramy$, and then the BLUE is necessarily equal to $W_o\vec{\hat{\total}}^{\mis}_\paramy$, a result that we were able to reproduce in our simulations.
This situation is particular to our simulation setup, that allows a small number of possible samples, but with a design for which the number of probable samples is larger than the number of month-in-sample estimates, the best linear estimator would not be exact. We computed the objective functions for $\alpha\in\{0,0.05,\ldots,1\}$ only. 
Table \ref{bestak} shows the optimal values for $a_1$, $k_1$, $a_2$, and $k_2$ for the three different populations and the best empirical estimator for level, change and compromise. The Census Bureau uses the coefficients $a_1=0.3$, $k_1=0.4$, $a_2=0.4$ and $k_2=0.7$ for the CPS. We notice that for each population, the best set of coefficients for change, level and compromise are very close, which means that the optimal choice for level is also almost optimal for change for those three populations.
        \begin{table}
        \centering\caption{Optimal $(a_1,k_1)$ and $(a_2,k_2)$ values for the three populations}\label{bestak}

        \begin{tabular}{@{}rrrr@{}}
        \toprule
          & Population 1 & Population 2 & Population 3\\ \midrule 
         $(a_1,k_1)$ (unemployed)\\Level & $(0.0471,0.85)$ & $(0.0395,0.398)$ & $(-0.0704,-0.619)$\\Compromise & $(0.029,0.895)$ & $(0.00175,0.0551)$ & $(0.0038,0.0253)$\\Change & $(0.0243,0.89)$ & $(0.0358,0.362)$ & $(-0.0239,-0.445)$\\\midrule $(a_2,k_2)$ (employed)\\Level & $(0.0714,0.752)$ & $(0.0453,0.73)$ & $(-0.0354,0.825)$\\Compromise & $(-0.0075,-0.232)$ & $(0.002,0.0598)$ & $(0.0464,0.0482)$\\Change & $(-0.0187,-0.256)$ & $(0.0658,0.723)$ & $(-0.0529,0.836)$\\ \bottomrule
        \end{tabular}
        
        \end{table}
Table \ref{bestrc}  shows  the best coefficient $\alpha$ for the regression composite estimators. 
\setlength{\tabcolsep}{5pt}
 \begin{table}[h]         
        \centering\caption{Optimal regression composite estimator's $\alpha$ parameter value for  three different populations }\label{bestrc}
        \begin{tabular}{rrrr}
        \toprule
    &Population 1\  & Population 2\  & Population 3\\
         \midrule 
         Level & $0.55$ ($0.6$) & $0.45$ ($0.6$) & $0$ \\Change & $1$  & $0.75$  & $0.8$ \\Compromise & $0.55$ ($0.6$) & $0.45$ ($0.6$) & $0$ \\
        \bottomrule
        \end{tabular}
        
        {\footnotesize Numbers in the  parentheses indicate parameter values in presence of rotation with bias when different}

        \end{table}

\subsection{Analysis without measurement error}

\begin{figure}[h]
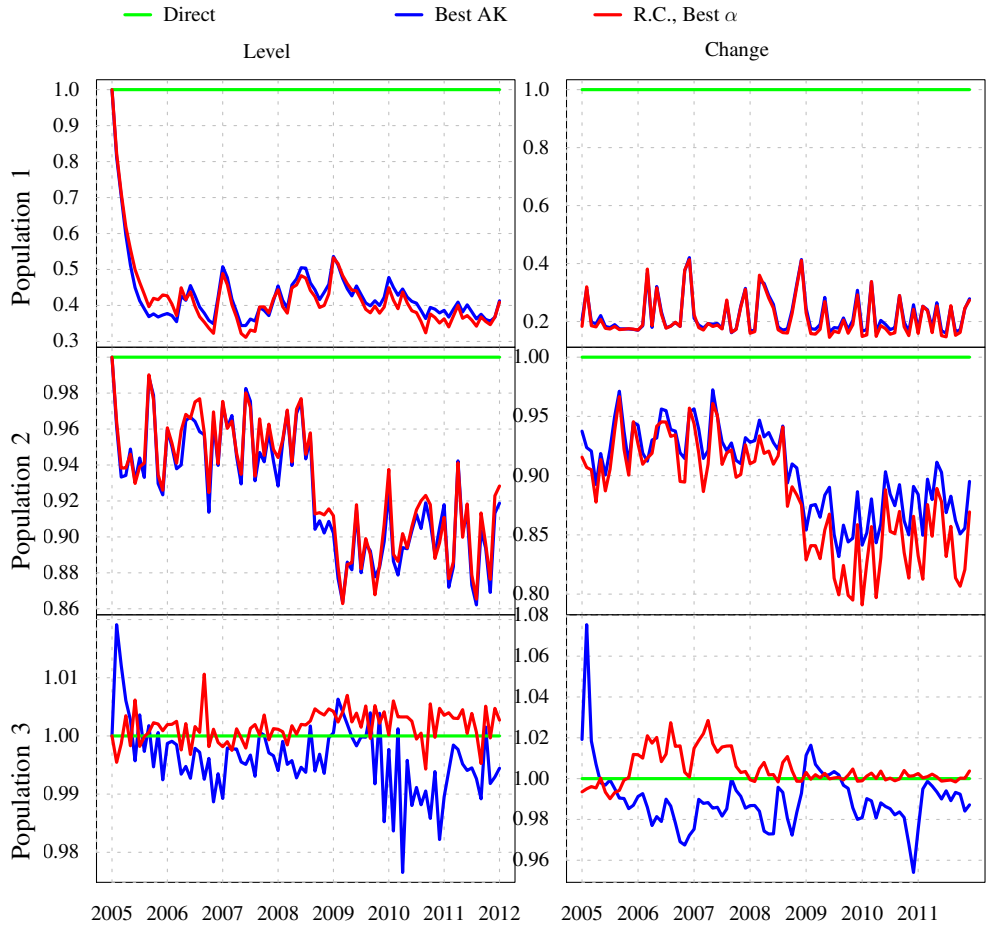

\begin{centering}
\caption{Relative mean squared errors of different estimated series of unemployment level and of month-to-month changes}\label{fig:1}



\end{centering}
 \end{figure}

Figure \ref{fig:1} displays the relative mean squared error for the different estimators of unemployment level and change, i.e. the times series :
$\left(\frac{\mathrm{MSE}\left[\hat{\ur}^\star_\mon\right]}{\mathrm{MSE}\left[\hat{\ur}^\direct_\mon\right]}\right)_{\mon\in\{1,\ldots,\Mon\}}$, and

\noindent $\left(\frac{\mathrm{MSE}\left[\hat{\ur}^\star_\mon-\hat{\ur}^\star_{\mon-1}\right]}{\mathrm{MSE}\left[\hat{\ur}^\direct_\mon-\hat{\ur}^\direct_{\mon-1}\right]}\right)_{\mon\in\{2,\ldots,\Mon\}},$ for $\star\in \{\direct,\ak,\rcind\}.$
In this figure, the best representative in each class is chosen, in the sense that the coefficients of Tables \ref{bestak} and \ref{bestrc} are used.
Note that in the absence of measurement error, the performances of all best ``estimators'' are comparable.

When trying to estimate the best A and K, then the results differ:  Table \ref{tab:3} (resp. \ref{tab:4}) represent the quantiles of the relative mean squared errors for different populations and 
for the best AK estimator, the empirical best AK estimator, the AK estimator with coefficient taken arbitrarily equal to the CPS AK coefficients (Arb. AK column), the best 
regression composite estimator (\rc column) and the Regression Composite estimator with $\alpha$ taken arbitrarily equal to $0.75$ (Arb. AK column) for the level (resp. change) estimation. 
Then for all  population, the arbitrary regression composite estimator seems to behave much better than the estimated best AK estimator, and arbitrary estimators, that perform worse than the direct.
The estimation of the best linear estimator gives even worse results than the estimated best AK and is not reported.
This underlines the weakness of the AK and Yansaneh Fuller-type estimators: without a good estimator of the variance matrix, they perform very poorly. Although the regression composite estimator with arbitrary $\alpha$ performs better without requiring any estimation of the variance.

{
\scriptsize
\setlength{\tabcolsep}{3pt}
\begin{table*}
\centering
\caption{Quantiles and means (over months) of the relative mean squared errors for different population and unemployment level estimators}\label{tab:3}
\scriptsize
\begin{tabular}{@{}rrrrrrcrrrrrcrrrrr@{}}\toprule
&
\multicolumn{5}{c}{Population 1} &
\phantom{a}&
\multicolumn{5}{c}{Population 2} &
\phantom{a} &
\multicolumn{5}{c}{Population 3}
\\
\cmidrule{2-6}
\cmidrule{8-12}
\cmidrule{14-18}
& Best & Arb. &Emp. & Best & Arb. &
& Best & Arb. &Emp. & Best & Arb. &
& Best & Arb. &Emp. & Best & Arb. \\
& AK &AK  &AK  & \rc &\rc&
& AK &AK  &AK& \rc &\rc  &
& AK &  AK&AK& \rc &\rc 

\\
\midrule
 0\% & $0.318$ & $1$ & $1$ & $0.322$ & $0.477$ &  & $0.87$ & $1$ & $1$ & $0.863$ & $0.885$ &  & $0.983$ & $1$ & $1$ & $0.994$ & $0.994$\\25\% & $0.377$ & $1.52$ & $2.59$ & $0.38$ & $0.546$ &  & $0.906$ & $1.35$ & $2.56$ & $0.913$ & $0.94$ &  & $0.996$ & $1.08$ & $1.03$ & $1$ & $1.01$\\50\% & $0.409$ & $1.6$ & $2.64$ & $0.42$ & $0.591$ &  & $0.929$ & $1.41$ & $2.7$ & $0.945$ & $0.974$ &  & $0.997$ & $1.14$ & $1.04$ & $1$ & $1.02$\\75\% & $0.454$ & $1.95$ & $2.74$ & $0.472$ & $0.663$ &  & $0.951$ & $1.49$ & $2.79$ & $0.969$ & $0.989$ &  & $1$ & $1.26$ & $1.07$ & $1$ & $1.02$\\100\% & $1$ & $2.09$ & $2.86$ & $1$ & $1$ &  & $1$ & $1.68$ & $3.08$ & $1$ & $1.02$ &  & $1.01$ & $1.65$ & $1.14$ & $1.01$ & $1.15$\\Mean & $0.431$ & $1.72$ & $2.64$ & $0.443$ & $0.613$ &  & $0.926$ & $1.42$ & $2.66$ & $0.94$ & $0.966$ &  & $0.997$ & $1.19$ & $1.05$ & $1$ & $1.02$
\\
\bottomrule
\end{tabular}

\end{table*}
}
\normalsize
\FloatBarrier
\setlength{\tabcolsep}{2.5pt}

{\scriptsize
\begin{table*}
\centering\caption{Quantiles and means (over months) of the relative mean squared errors for different populations and unemployment month-to-month change estimators}\label{tab:4}

\scriptsize
\begin{tabular}{@{}rrrrrrcrrrrrcrrrrr@{}}\toprule
&
\multicolumn{5}{c}{Population 1} &
\phantom{a}&
\multicolumn{5}{c}{Population 2} &
\phantom{a} &
\multicolumn{5}{c}{Population 3}
\\
\cmidrule{2-6}
\cmidrule{8-12}
\cmidrule{14-18}
& Best & Arb. &Emp. & Best & Arb. &
& Best & Arb. &Emp. & Best & Arb. &
& Best & Arb. &Emp. & Best & Arb. \\
& AK &AK  &AK  & \rc &\rc&
& AK &AK  &AK& \rc &\rc  &
& AK &  AK&AK& \rc &\rc

\\
\midrule
0\% & $0.0959$ & $2.77$ & $5.43$ & $0.0279$ & $0.0936$ &  & $0.845$ & $0.872$ & $2.72$ & $0.774$ & $0.791$ &  & $0.973$ & $0.998$ & $1.01$ & $0.984$ & $0.994$\\25\% & $0.123$ & $3.31$ & $6.35$ & $0.0455$ & $0.112$ &  & $0.887$ & $0.953$ & $3.07$ & $0.835$ & $0.847$ &  & $0.99$ & $1.02$ & $1.03$ & $0.992$ & $1$\\50\% & $0.142$ & $3.68$ & $6.64$ & $0.0552$ & $0.127$ &  & $0.914$ & $0.998$ & $3.33$ & $0.885$ & $0.89$ &  & $0.993$ & $1.02$ & $1.04$ & $0.997$ & $1$\\75\% & $0.215$ & $5.21$ & $6.93$ & $0.146$ & $0.201$ &  & $0.932$ & $1.03$ & $3.62$ & $0.916$ & $0.919$ &  & $0.996$ & $1.03$ & $1.06$ & $1$ & $1$\\100\% & $0.395$ & $6.12$ & $7.59$ & $0.355$ & $0.383$ &  & $0.971$ & $1.13$ & $3.92$ & $0.965$ & $0.967$ &  & $1.04$ & $1.06$ & $1.14$ & $1.11$ & $1.01$\\Mean & $0.174$ & $4.21$ & $6.68$ & $0.102$ & $0.163$ &  & $0.909$ & $0.993$ & $3.33$ & $0.876$ & $0.883$ &  & $0.993$ & $1.03$ & $1.04$ & $1$ & $1$
\\
\bottomrule
\end{tabular}
\end{table*}

}
\normalsize

\subsection{Analysis with measurement error}

Under \eqref{M1}, a solution to the rotation group bias for adapting the AK estimator consists in estimating the rotation bias parameter vector $\rotationbias$ and in applying AK coefficients to corrected month-in-sample estimates, to obtain $\left(\hat{\total}^{\ak*}_\paramy\right)_{\mon,.}=\sum_{\mon'=1}^\mon\sum_{\mon'=1}^\mon\left( c_{\mon,\mon',\misi} \left(\hat{\total}^{\mis,\misi}_\paramy\right)_{\mon,\misi,.}-\hat{\rotationbias}_\misi \right).$
The question of how to adapt the regression composite estimator to take into account measurement error is more complicated. Besides, the model used for rotation bias is itself questionable. The  linear constraint on $\rotationbias$ ($\sum \rotationbias_{\misi,.} =0$ or $\rotationbias_{1,.}=0$) is imposed to address an identifiability problem, but one cannot assess its validity. This is why we think it is not a good way to deal with the rotation bias. We  have not investigated  how to adapt the regression composite estimator to address the problem of rotation bias (we think that rotation bias has to be studied at the individual level throught resampling method). Instead we studied its behaviour  in presence of rotation bias.
To this end, we systematically (for all month, all sample) changed up to 2 unemployed persons of month-in-sample  group 1 status from unemployed to employed.
Tables \ref{tab:6} and \ref{tab:7} display for different populations the quantiles and means of the relative mean square errors for the level and change estimation and for the best AK estimator and the best regression composite estimator.
We applied  the best AK and best regression composite estimators for the case without measurement error to the case with measurement error.
We notice that AK estimator is very sensitive to rotation bias, whereas regression composite estimator is not. A reason may be that introducing a variable not correlated to the study variables in the calibration procedure does not much change the estimation of the study variable. Rotation bias weakens the correlation between $\paramz$ and $\paramy$ and though the performance of the regression composite estimator is comparable to the performance of the direct.

{
\scriptsize
\setlength{\tabcolsep}{4pt}
\begin{table*}
\centering
\caption{Quantiles and means (over months) of the relative mean squared errors for different population and unemployment level estimators}\label{tab:6}
\scriptsize
\begin{tabular}{@{}rrrcrrcrrcrrcrrcrr@{}}\toprule
&
\multicolumn{2}{c}{Population 1} &
\phantom{a}&
\multicolumn{2}{c}{Population 2} &
\phantom{a} &
\multicolumn{2}{c}{Population 3} &
\phantom{a} &
\multicolumn{2}{c}{Pop. 1 (bias)} &
\phantom{a}&
\multicolumn{2}{c}{Pop. 2 (bias)} &
\phantom{a} &
\multicolumn{2}{c}{Pop. 3  (bias)}
\\
\cmidrule{2-3}
\cmidrule{5-6}
\cmidrule{8-9}
\cmidrule{11-12}
\cmidrule{14-15}
\cmidrule{17-18}
& AK & \rc & & AK & \rc & & AK & \rc && AK & \rc & & AK & \rc && AK & \rc
\\
\midrule
 0\% & $0.318$ & $0.322$ &  & $0.87$ & $0.863$ &  & $0.983$ & $0.994$ &  & $1$ & $0.0521$ &  & $1$ & $0.117$ &  & $0.919$ & $0.158$\\25\% & $0.377$ & $0.38$ &  & $0.906$ & $0.913$ &  & $0.996$ & $1$ &  & $46.1$ & $0.0735$ &  & $2.78$ & $0.155$ &  & $1.5$ & $0.739$\\50\% & $0.409$ & $0.42$ &  & $0.929$ & $0.945$ &  & $0.997$ & $1$ &  & $47.9$ & $0.0949$ &  & $2.81$ & $0.179$ &  & $1.59$ & $0.768$\\75\% & $0.454$ & $0.472$ &  & $0.951$ & $0.969$ &  & $1$ & $1$ &  & $52.5$ & $0.115$ &  & $2.86$ & $0.254$ &  & $1.86$ & $0.786$\\100\% & $1$ & $1$ &  & $1$ & $1$ &  & $1.01$ & $1.01$ &  & $57.6$ & $0.162$ &  & $2.92$ & $0.3$ &  & $2.18$ & $0.843$\\Mean & $0.431$ & $0.443$ &  & $0.926$ & $0.94$ &  & $0.997$ & $1$ &  & $45.6$ & $0.0957$ &  & $2.77$ & $0.203$ &  & $1.64$ & $0.754$
\\
\bottomrule
\end{tabular}
\end{table*}
}
\normalsize
\FloatBarrier
\setlength{\tabcolsep}{3.5pt}

{\scriptsize
\begin{table*}
\centering
\caption{Quantile and means (over months) of the relative mean squared errors for different population and unemployment month-to-month change estimators}\label{tab:7}
\scriptsize
\begin{tabular}{@{}rrrcrrcrrcrrcrrcrr@{}}\toprule
&
\multicolumn{2}{c}{Population 1} &
\phantom{a}&
\multicolumn{2}{c}{Population 2} &
\phantom{a} &
\multicolumn{2}{c}{Population 3} &
\phantom{a} &
\multicolumn{2}{c}{Pop. 1 (bias)} &
\phantom{a}&
\multicolumn{2}{c}{Pop. 2 (bias)} &
\phantom{a} &
\multicolumn{2}{c}{Pop. 3  (bias)}
\\
\cmidrule{2-3}
\cmidrule{5-6}
\cmidrule{8-9}
\cmidrule{11-12}
\cmidrule{14-15}
\cmidrule{17-18}
& AK & \rc & & AK & \rc & & AK & \rc && AK & \rc & & AK & \rc && AK & \rc
\\
\midrule
 0\% & $0.0959$ & $0.0936$ &  & $0.845$ & $0.791$ &  & $0.973$ & $0.994$ &  & $0.422$ & $0.0298$ &  & $0.898$ & $0.457$ &  & $1.19$ & $0.935$\\25\% & $0.123$ & $0.112$ &  & $0.887$ & $0.847$ &  & $0.99$ & $1$ &  & $0.477$ & $0.0385$ &  & $0.938$ & $0.552$ &  & $1.48$ & $0.994$\\50\% & $0.142$ & $0.127$ &  & $0.914$ & $0.89$ &  & $0.993$ & $1$ &  & $0.515$ & $0.05$ &  & $0.954$ & $0.583$ &  & $1.5$ & $1.01$\\75\% & $0.215$ & $0.201$ &  & $0.932$ & $0.919$ &  & $0.996$ & $1$ &  & $0.563$ & $0.093$ &  & $0.971$ & $0.613$ &  & $1.52$ & $1.02$\\100\% & $0.395$ & $0.383$ &  & $0.971$ & $0.967$ &  & $1.04$ & $1.01$ &  & $3.96$ & $0.209$ &  & $1.25$ & $0.673$ &  & $1.58$ & $1.05$\\Mean & $0.174$ & $0.163$ &  & $0.909$ & $0.883$ &  & $0.993$ & $1$ &  & $0.665$ & $0.0671$ &  & $0.958$ & $0.581$ &  & $1.48$ & $1$
\\
\bottomrule
\end{tabular}
\end{table*}
}
\normalsize

\section{The CPS Data Analysis}
\subsection{Implementation of regression composite estimator for the CPS}
\subsubsection{Choice of $\alpha$}\label{sec:5.2.2}

Under a simple unit level  times series model  with auto-regression coefficient $\rho$, Fuller and Rao (2001) proposed a formal expression for an approximately optimal $\alpha$ as a function of $\rho$ and studied the so-called drift problem for the MR2 choice: $\alpha=1$.  They also proposed approximate expressions for variances of their estimators for the level and change.
For various reasons, it seems difficult to obtain the optimal or even an approximately optimal $\alpha$ needed for the Fuller-Rao type regression composite estimation technique to produce the U.S. employment and unemployment rates using the CPS data.  First of all, the simple time series model used by Fuller and Rao (2001) is not suitable to model a nominal variable (employment status) with several categories.  Secondly, complexity of the CPS design poses a challenging modeling problem. Before attempting to obtain the optimal or even an approximately optimal choice for $\alpha$ required for  the Fuller-Rao type regression composite method, it will be instructive to evaluate the regression composite estimators for different known choices of $\alpha$.  This is the focus of this section.

\subsubsection{Choice of $\paramx$ and $\paramz$}\label{sec:5.1.2}

In our study, we considered two candidates for $\paramz$: (i) $\paramz=\paramy$, (ii) a more detailed employment status variable with 8 categories. As the use of this variable reduces the degrees of freedom in the calibration procedure and leads to estimates with a higher mean square error, we just report on our first choice.
 For an application of the Fuller-Rao method, 
 one might think of including all the variables that have been already used for the weight adjustments in the $\paramx$ variables.  However, this would introduce many constraints on the coefficients and thus is likely to cause a high variability in the ratio of
$\paramw_{\mon,\indiv}$ and  $\paramw_{\mon,\indiv}^{\rcind}$. The other extreme option is not to use any of these auxiliary variables.  But then the final weights  would not be adjusted for the known totals of auxiliary variables $\paramx$.
As a compromise, we selected only two variables: gender and race.

\subsection{Results}

\begin{figure}[h]
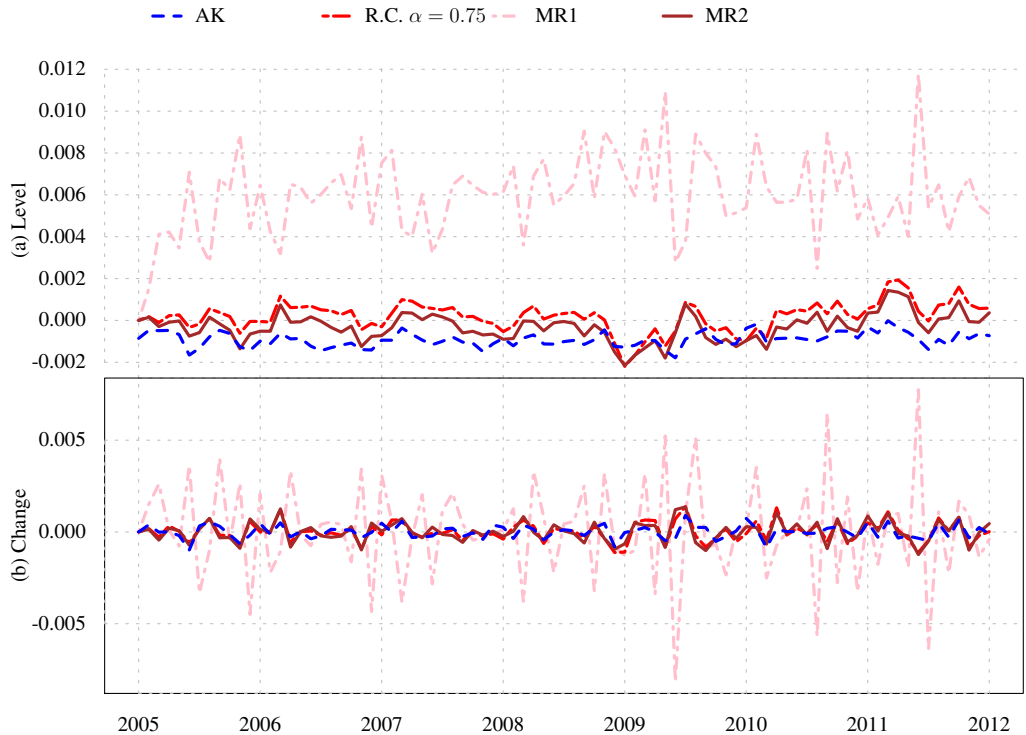

\begin{center}
\caption{Estimated series of differences between different composite estimates and the corresponding direct estimates }\label{fig:2}
\hspace*{-1cm}


\end{center}
\end{figure}

Figure \ref{fig:2}(a) displays 
the difference $\widehat{\rate}^{\ak}_\mon-\widehat{\rate}^{\direct}_\mon$ between different composite estimates and the corresponding direct estimates against months $\mon$.  For the regression composite estimator, we considered three choices: (i)  $\alpha=0.75$ (suggested by Fuller and Rao), (ii) $\alpha=0$ (corresponding to MR1), and (iii) $\alpha=1$ (corresponding to MR2).
We display similar graphs for month-to-month  change estimates in Figure \ref{fig:2}(b).  Notice that $\alpha=0$ and $\alpha=1$ correspond to MR1 and MR2, respectively.
We display similar graphs for month-to-month  change estimates in Figure \ref{fig:2}.

It is interesting to note that the AK composite estimates of unemployment rates are always lower than the corresponding direct estimates in Figure \ref{fig:2}(a).  To our knowledge, this behavior of AK composite estimates has not been noticed earlier.  In contrast, the regression composite estimates MR1 are always higher than the corresponding direct estimates.  However, such deviations  decrease as $\alpha$ gets closer to 1 in Figure \ref{fig:2}(a).  Application of the Fuller-Rao method at the household level causes an increase in the distance between the original and calibrated weights and one may expect an increase in  the variances of the estimates.  Figure \ref{fig:2}(b) does not indicate systematic deviations of the composite estimates of the month-to-month changes from the corresponding direct estimates.  Deviations of the regression composite estimates from the corresponding direct estimates seem to decrease as $\alpha$ approaches  1.

\section{Discussion}

Our study reveals that there is ample scope for improving the AK estimator used by the Census Bureau.  
We would like to emphasize the following  undesirable features of the AK estimation method:

(i)
the method used to compute optimal coefficient is crude --- the best coefficients are just picked from 10 different values. Our R package, based on the built in R Nelder-Mead algorithm, can provide the optimal coefficients within 8 digits of precision in a reasonable time.  

(ii) 
The stationarity assumption on the variances and covariances of the month-in-sample estimators over a period of 10 years does not seem realistic, and to our knowledge, has not been tested before. Besides, even though the stationary model was reasonable, the complexity of the CPS design makes it difficult to evaluate the quality of the estimators used for that model. The difficulty to propose a stochastic model in advance for the best linear estimators in the CPS was already pointed out earlier by \cite[Sec.~4]{jones1980best}. Our evaluation study shows that the AK estimators is very sensitive to the choices of A and K, and that the errors in the estimation of the  variances and covariances may lead to poor performance of the AK estimators. We add that estimators of  variances and covariances of month-in-sample estimators estimation also affect the performances of  empirical best linear unbiased estimators. 
 
 (iii)
Using the Bailar model for the bias in our study, we showed that AK estimator is very sensible to rotation group bias. There is currently no satisfactory way to correct the AK estimator for the rotation bias. The Bailar model relies on an arbitrary constraint on the month-in-sample biases and a strong stationarity assumption of the month-in-sample bias and should not be used unless some re-interview study can justify the Bailar's model.  

 (iv)
 The computation of composite weights in CPS to calibrate the weights on the AK estimators will affect all other weighted estimators.
Although \cite{Lent1996} showed that there was not a big effect on the estimates, considering the concerns about AK estimators listed before, we do not think that the use of  those composite weights is a good option.

 (v)
the CPS data analysis shows that the AK estimates are consistently smaller than the corresponding direct survey-weighted estimates for the period 2005-2012. This is also a source of concern. 
 
The composite regression estimator does not rely on an estimation of the variances and covariances matrix. In our simulation study, it appears to be less sensitive to rotation group bias, and bounces around the survey-weighted estimates when applied to the real CPS data. Our study encourages the use of the regression composite method in the US labor force estimation.

To facilitate and encourage further research on this important topic, we make the following three R packages, developed under this project, freely available: (i) the package dataCPS downloads CPS public data files and transform them into R data set (\cite{github:dataCPS}); (ii) the package CompositeRegressionEstimation allows computation of the AK, best AK, composite regression, linear and best linear estimators (\cite{github:CompositeRegressionEstimation}); (iii) the  package pubBonneryChengLahiri2016 allows to reproducing all computations and simulations of this paper \citep{github:pubBonneryChengLahiri2016}.

\bibliographystyle{apalike}

\section*{Acknowledgements}
The research of the first and third authors has been supported by the U.S. Census Bureau Prime Contract No: YA1323-09-CQ-0054 (Subcontract No: 41-1016588).
{The programs used for the simulations have been made available on the github repository \cite{github:pubBonneryChengLahiri2016}}.

\appendix

{
\section{Description of CPS design}\label{ap:cps}
This section uses CPS notations for rotation groups.
      Let $U$ be the intersection of a given basic
primary sampling unit  component (BPC) and one of the frames used in CPS (see \cite{CPS2006}). The BPC is a set of clusters of about four housing units, the clusters are the ultimate sampling units (USU).
Let $N$ be the number of clusters in $U$.
The clusters in $U$ are sorted according to geographical and demographic characteristics and then indexed by $k=1\ldots N$. In the sequence, we will designate a cluster by its index.
Let $SI_w$ be the adjusted within-PSU sampling interval, as defined in \cite[p.~3-11]{CPS2006}.
Let $n=\left\lfloor(21\times 8*SI_w)^{-1} N\right\rfloor$, where $\lfloor.\rfloor$ is the floor function. The number $n$ is the sample size for a sample rotation group.
The drawing of the USU within the PSU consists in the generation of a random number $X$ according to the uniform law on $[0,1]$.
For $i=1\ldots n$, $j=1\ldots 8$, $\ell=85\ldots (85+15)$, let $k_{i,j,\ell}$ denote  the cluster
$k_{i,j,\ell}=\lfloor (X+8\times (i-1)+j)\times SI_w+(\ell-85)\rfloor$.
Then, with the notations of \cite{CPS2006} for $\ell=85\ldots 100 $, $j=1\ldots 8$, the rotation group $j$ of sample $A_\ell$ is
$$A_{\ell,j}=\left\{k_{i,j,\ell}\mid i=1\ldots n\right\} .$$

For a given month the sample consits of 8 rotation groups.
There are 120 months in a period of 10 years.
For $m=1\ldots 120$, $j'\in \left\{1,\ldots,8\right\}$,
$\ell_{m,j'}$ and $j_{m,j'}$ are given by:
$j_{m,j'}=t+j'-1-8\times \left\lfloor(t+j'-2)/8\right\rfloor$.
If $j'\in\left\{1,\ldots,4\right\}$,
$\ell_{m,j'}=85+\left\lfloor{(t+j'-2)/8}\right\rfloor$.
If $j'\in\left\{5,\ldots,8\right\}$,
$\ell_{m,j'}=86+\left\lfloor(t+j'-2)/8\right\rfloor$.

The sample of the $m$th month, counting from November 2009, is
$$s_m=\bigcup_{j'=1}^8 A_{\ell_{m,j'},j_{m,j'}}.$$

For example, June 2013 corresponds to $m=44$, counting from November 2009.
Then \begin{align*}
\ell_{m,1}&=85+\left\lfloor43/8\right\rfloor=90&
j_{m,1}&=44-8\times\left\lfloor43/8\right\rfloor=4\\
\ell_{m,2}&=85+\left\lfloor44/8\right\rfloor=90&
j_{m,2}&=45-8\times\left\lfloor44/8\right\rfloor=5
       \\
\ell_{m,3}&=85+\left\lfloor45/8\right\rfloor=90&
j_{m,3}&=46-8\times\left\lfloor45/8\right\rfloor=6
       \\
\ell_{m,4}&=85+\left\lfloor46/8\right\rfloor=90&
j_{m,4}&=47-8\times\left\lfloor46/8\right\rfloor=7
       \\
\ell_{m,5}&=86+\left\lfloor47/8\right\rfloor=91&
j_{m,5}&=48-8\times\left\lfloor47/8\right\rfloor=8
       \\
\ell_{m,6}&=86+\left\lfloor48/8\right\rfloor=92&
j_{m,6}&=49-8\times\left\lfloor48/8\right\rfloor=1
       \\
\ell_{m,7}&=86+\left\lfloor49/8\right\rfloor=92&
j_{m,7}&=50-8\times\left\lfloor49/8\right\rfloor=2
       \\
\ell_{m,8}&=86+\left\lfloor50/8\right\rfloor=92&
j_{m,8}&=51-8\times\left\lfloor50/8\right\rfloor=3
     \end{align*}

We can check from the CPS rotation chart \cite[Fig. 3-1]{CPS2006} that the sample of June 2013 consists of the
4th, 5th, 6th, 7th rotation groups of A90, of the 8th rotation group of A91, and of the
1st, 2d and 3rd rotation groups of A92:
 $$S_{\text{June 2013}}=A_{90,4}\cup A_{90,5}\cup A_{90,6}\cup A_{90,7}\cup A_{91,8}\cup A_{92,1}\cup A_{92,2} \cup A_{92,3}.$$
}
 \printindex[notations]
\end{document}